\documentclass[5p,twocolumn,times]{elsarticle}
\usepackage[pdftex]{color}
\usepackage{lipsum}
\usepackage{lineno,hyperref}
\modulolinenumbers[5]
\usepackage[pdftex]{color}
\usepackage[font=footnotesize,labelfont=bf]{caption}
\usepackage[font=footnotesize,labelfont=bf]{subcaption}
\journal{Journal of \LaTeX\ Templates}

\usepackage{amssymb}

\biboptions{compress}

\usepackage[figuresright]{rotating}

\begin{document}

\begin{frontmatter}

\title{Combination of survival movement strategies in cyclic game systems during an epidemic}

\address[1]{Escola de Ci\^encias e Tecnologia, Universidade Federal do Rio Grande do Norte\\
Caixa Postal 1524, 59072-970, Natal, RN, Brazil}

\address[2]{Departamento de Engenharia Biomédica, Universidade Federal do Rio Grande do Norte\\
  Av. Senador Salgado Filho, 300 Lagoa Nova,
CEP 59078-970 - Natal, RN - Brazil}
\address[3]{Edmond and Lily Safra International Neuroscience Institute, Santos Dumont Institute\\
Av Santos Dumont, 1560, 59280-000, Macaiba, RN, Brazil} 
\address[4]{Institute for Biodiversity and Ecosystem
Dynamics, University of Amsterdam, Science Park 904, 1098 XH
Amsterdam, The Netherlands}

\author[1]{E. Rangel} 
\author[2,3]{B. Moura} 
\author[1,4]{J. Menezes}

\begin{abstract}
Disease outbreaks affect many ecosystems threatening species that also fight against other natural enemies. 
We investigate a cyclic game system with $5$ species, whose organisms outcompete according to the rules of a generalised spatial rock-paper-scissors game, during an epidemic. We study the effects of behavioural movement strategies that allow individuals of one out of the species to move towards areas with a low density of disease vectors and a high concentration of enemies of their enemies. We perform a series of stochastic simulations to discover the impact of self-preservation strategies in pattern formation, calculating the species' spatial autocorrelation functions.
Considering organisms with different physical and cognitive abilities, we compute the benefits of each movement tactic to reduce selection and infection risks. 
Our findings show that the maximum profit in terms of territorial dominance in the cyclic game is achieved if both survival movement strategies are combined, with individuals prioritising social distancing. In the case of an epidemic causing symptomatic illness, the drop in infection risk when organisms identify and avoid disease vectors does not render a rise in the species population because many refuges are disregarded, limiting the benefits of safeguarding against natural enemies. Our results may be helpful to the understanding of the behavioural strategies in ecosystems where organisms adapt to face living conditions changes.
\end{abstract}



\end{frontmatter}

\section{Introduction}
\label{sec:int}

A central goal of ecology is to comprehend how interactions among organisms influence the formation and stability of ecosystems \cite{ecology,foraging,BUCHHOLZ2007401}. Following environmental clues, many animals behave strategically, adapting to the variations in their living conditions, facing more effectively the threats to their survival
\cite{butterfly,adaptive1,adaptive2,Dispersal,BENHAMOU1989375,Causes,MovementProfitable}. 
Besides the problematic survival conditions due to the competition for natural resources with other species, many ecosystems are affected by epidemics, whose disease spreading increases the risk of species going extinct \cite{Directional1,epidemicbook,epidemicprocess,COVID,tanimoto}. This leads organisms to adapt, combining survival strategies to prevail in the challenging scenario. For example, organisms use sensory information to detect and flee enemies, forming partnerships with other species and taking distance from disease vectors \cite{Odour,socialdist,soc}.
It has been shown that disease can mediate coexistence in ecosystems where individuals carry pathogens that determine the success of the invasion \cite{disease4,disease3,disease2}. If more than one disease is present, coexistence is more probable if their virulence and transmission are substantially different \cite{virulence}.

Many authors have demonstrated that organisms' spatial interactions are 
crucial to ecosystems stability \cite{Nature-bio}. It has been found that the cyclic dominance among three strains of bacteria \textit{Escherichia coli} can be described by the RPS game rules; however, the experiments revealed that the coexistence holds only if organisms interact locally \cite{Coli,bacteria,Allelopathy}.
For this reason, cyclic games systems have been widely studied using stochastic simulations of the spatial rock-paper-scissors game \cite{Reichenbach-N-448-1046,Moura, Anti1,anti2,Avelino-PRE-86-036112,park1}. In this class of models, organisms move randomly or directionally according to behavioural strategies motivated by environmental attack or defence
stimuli \cite{Moura,PhysRevE.97.032415,MENEZES2022101606}. 
The stability of species populations in cyclic models has also been investigated by evolutionary pairwise Fermi rules, 
where both linear and nonlinear dynamics have been demonstrated to yield similar equilibrium conditions \cite{Cheng-SR-4-7486,nowaknew,tanimoto2}.
Therefore, both cyclic dominance among species and the presence of a disease may be crucial to promote biodiversity, as happens in lizard communities
 \cite{lizards,disease8}.
 
This work
investigates a cyclic model composed of five species whose organism faces invasion of organisms of a dominant species and an infectious disease \cite{rps-epidemy,epidemic-graphs,germen}. All organisms of every species are equally susceptible to being contaminated; once sick, the organisms may die because of the disease complications or be cured - being subjected to reinfection.
Our goal is to understand how survival strategies can be combined to result in population growth for the species that move strategically. We simulate two self-preservation movement tactics: i) safeguard against death by competition: organisms approach their enemies' enemies; ii) self-protection against disease infection: organisms moving towards less populated areas, where the density of empty spaces is high.
We run a series of stochastic simulations considering organisms with
different physical and cognitive abilities to perform the collective strategy. We also compute the influence of the movement strategies in the selection and infection risks in epidemic scenarios where a fraction of sick organisms are distinguished because of the symptoms. 

The outline of this paper is as follows. In Sec.~\ref{sec2}, we describe the Methods, introducing the stochastic model, the survival movement strategies, the simulation implementation, and the model parameters. In Sec.~\ref{sec3}, we investigate how each survival movement tactic impacts the pattern formation process and population dynamics. In Sec.~\ref{sec4}, we compute the species autocorrelation function and the characteristic length of the typical areas occupied by each species in terms of the organisms' perception radius. The dependence of the selection and infection risks as well the species population densities on the organism's conditioning factor is studied in Sec.~\ref{sec5}. In Sec. ~\ref{sec6}, the combination of survival movement strategies is addressed, with the conditions to maximise the benefits in terms of territorial control being found. The influence of the development of symptomatic disease by infected organisms in the results of the combined survival strategies is investigated in Sec.~\ref{sec7}. Finally, the outcomes are discussed, and conclusions are highlighted in Sec.~\ref{sec8}.


\section{Methods}
\label{sec2}

\subsection{The model}

We study a cyclic game system whose selection dominance can be described by the generalised rock-paper-scissors model with $5$ species. Organisms competing each other, with the dominance according to the arrows in the illustration in Fig. 1, with species $i$ beating species $i+1$, with $i=1,2,3,4,5$, with the identification $i=i+5\,\kappa$, where $\kappa$ is an integer. We consider that an epidemic outbreak affects all species, being transmitted person-to-person with the same probability. Once infected, individuals may either die due to the illness severity or be cured, being subject to reinfection. 
We investigate two survival movement strategies performed by organisms of one of the species: i) safeguard from being eliminated by the dominant species; ii) self-protection from disease infection.
Considering that an infected organism of any species can transmit the disease to one another and that all organisms are equally susceptible to being contaminated, we study two epidemic scenarios: asymptomatic or symptomatic disease. In the latter case, neighbours can identify infected organisms, while they are assumed to be healthy in the former because of the lack of symptoms.

The dynamics of individuals' spatial organisation occurs in square lattices with periodic boundary conditions. We assume the May-Leonard implementation, which means that the total number of individuals is not conserved \cite{leonard}. Each grid point contains at most one individual; thus, the maximum number of individuals is $\mathcal{N}$, the total number of grid points. 
Initially, the number of individuals is the same for all species, i.e., $I_i\,=\,\mathcal{N}/5$, with $i=1,2,3,4,5$ (there are no empty spaces in the initial state). We prepared the initial conditions by distributing each individual at a random grid point. At each timestep, one interaction occurs, changing the spatial configuration.

Let us define $h_i$ and $s_i$ to identify healthy and sick individuals of species $i$; while $i$ stands for individuals irrespective of illness or health. With this notation, we describe the interactions as follows:
\begin{itemize}
\item 
Selection: $ i\ j \to i\ \otimes\,$, with $ j = i+1$, where $\otimes$ means an empty space; every time one selection interaction occurs, the grid point occupied by the individual of species $i+1$ becomes empty.
\item
Reproduction: $ i\ \otimes \to i\ i\,$; a new individual of species $i$ is created, filling an empty space.
\item 
Mobility: $ i\ \odot \to \odot\ i\,$, where $\odot$ means either an individual of any species or an empty site; an individual of species $i$ switches grid site with another organism of any species or with an empty space.
\item 
Infection: $ s_i\ h_j \to s_i\ s_j\,$, with $j=1,2,3,4,5$; a sick individual of species $i$ infect a healthy individual of any species.
\item 
Cure: $ s_i \to h_i\,$; a sick individual of species $i$ is naturally cured of the disease.
\item 
Death: $ s_i \to \otimes\,$; a sick individual of species $i$ dies due to the disease, leaving its position empty.
\end{itemize}

In our simulations, interactions are implemented according to the set of probabilities: i) heathy individuals: $s_h$, $r_h$, $m_h$ for selection, reproduction, and mobility, respectively;
ii) sick individuals: $s_s$, $r_s$, $m_s$, $w$, $c$, $d$,  for selection, reproduction, mobility, infection, cure, and death respectively. The probabilities are the same for all organisms of every species.
The interactions were implemented by assuming the von Neumann neighbourhood, i.e., individuals may interact with one of their four immediate neighbours. The simulation algorithm follows three steps: i) randomly choosing an active individual; ii) raffling one interaction to be executed; iii) drawing one of the four nearest neighbours to suffer the interaction - the only exception is the directional mobility, where the neighbour is not random, by chosen according to the survival strategy performed by the active individual. If the interaction is executed, one timestep is counted. Otherwise, the three steps are repeated. Our time unit is called generation, defined as the necessary time to $\mathcal{N}$ timesteps to occur. The densities of species $i$ at time $t$ is defined as 
$\rho_i = I_i/\mathcal{N}$, with $i=1,2,3,4,5$.

\begin{figure}
\centering
\includegraphics[width=40mm]{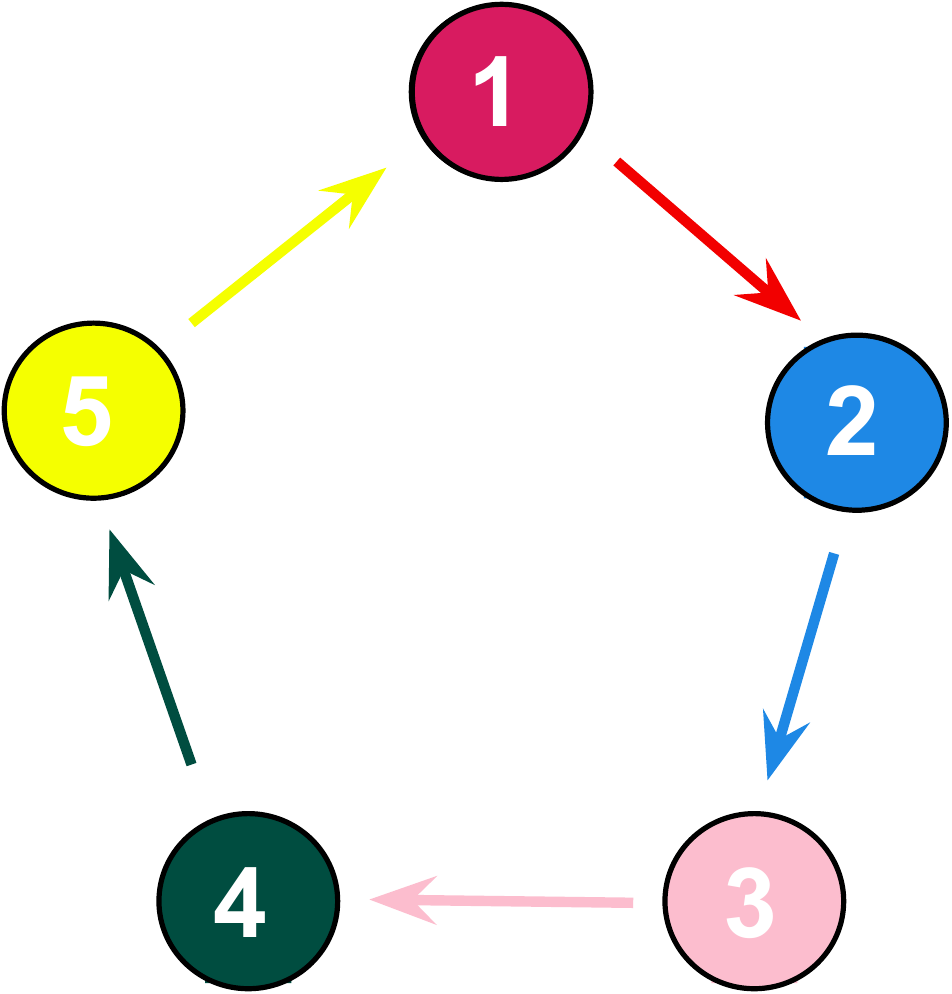}
\caption{Illustration of the cyclic selection interactions in the generalised rock-paper-scissors model with $5$ species. Arrows show the dominance of organisms of species $i$ over individuals of species $i+1$.}
\label{fig1}
\end{figure}
\begin{figure*}
\centering
    \begin{subfigure}{.22\textwidth}
        \centering
        \includegraphics[width=40mm]{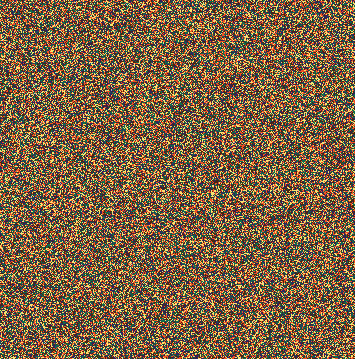}
        \caption{}\label{fig2a}
    \end{subfigure} %
    \begin{subfigure}{.22\textwidth}
        \centering
        \includegraphics[width=40mm]{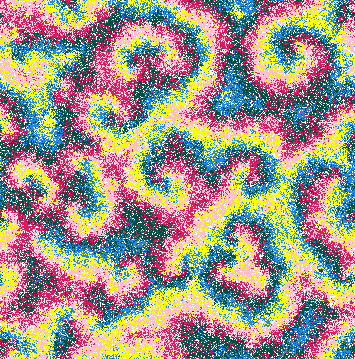}
        \caption{}\label{fig2b}
    \end{subfigure} %
       \begin{subfigure}{.22\textwidth}
        \centering
        \includegraphics[width=40mm]{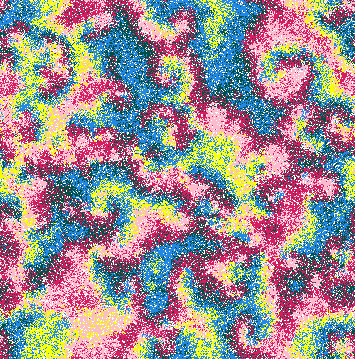}
        \caption{}\label{fig2c}
    \end{subfigure} %
           \begin{subfigure}{.22\textwidth}
        \centering
        \includegraphics[width=40mm]{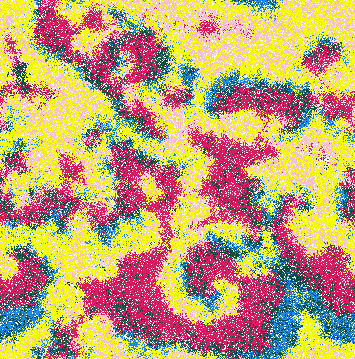}
        \caption{}\label{fig2d}
    \end{subfigure} %
\caption{Snapshots captured from simulations in lattices with $300^2$ grid points of the cyclic model described in Fig.~\ref{fig1}. Figure a shows the random initial conditions used in Simulation A (standard model), Simulation B (safeguard strategy), and Simulation C (social distancing), whose spatial configurations at $t=3000$ generations are shown in Figs. b and c. The colours follow the scheme in Fig.~\ref{fig1}.}
  \label{fig2}
\end{figure*}
\begin{figure*}
 \centering
        \begin{subfigure}{.33\textwidth}
        \centering
        \includegraphics[width=57mm]{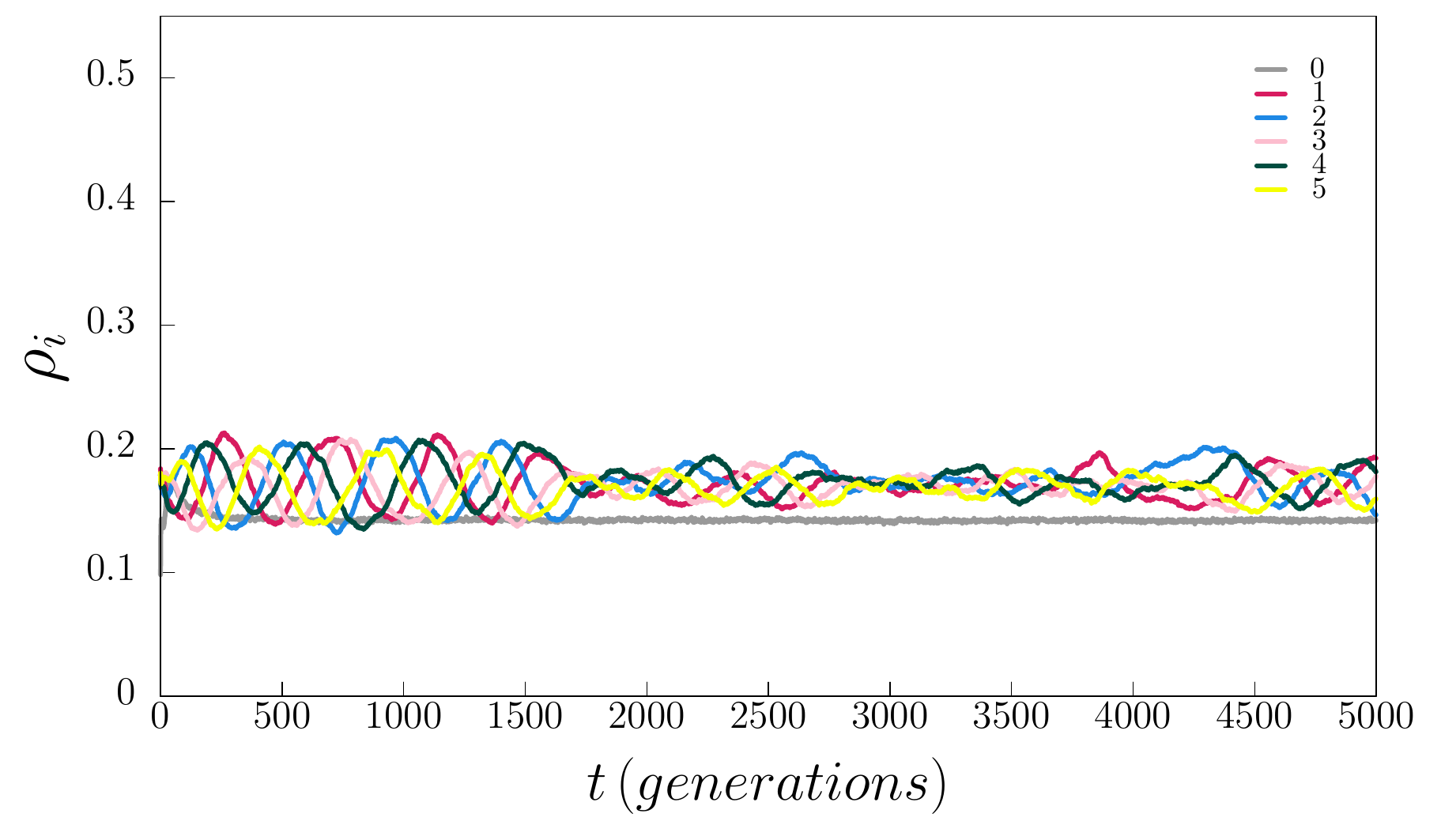}
        \caption{}\label{fig3a}
    \end{subfigure}
       \begin{subfigure}{.33\textwidth}
        \centering
        \includegraphics[width=57mm]{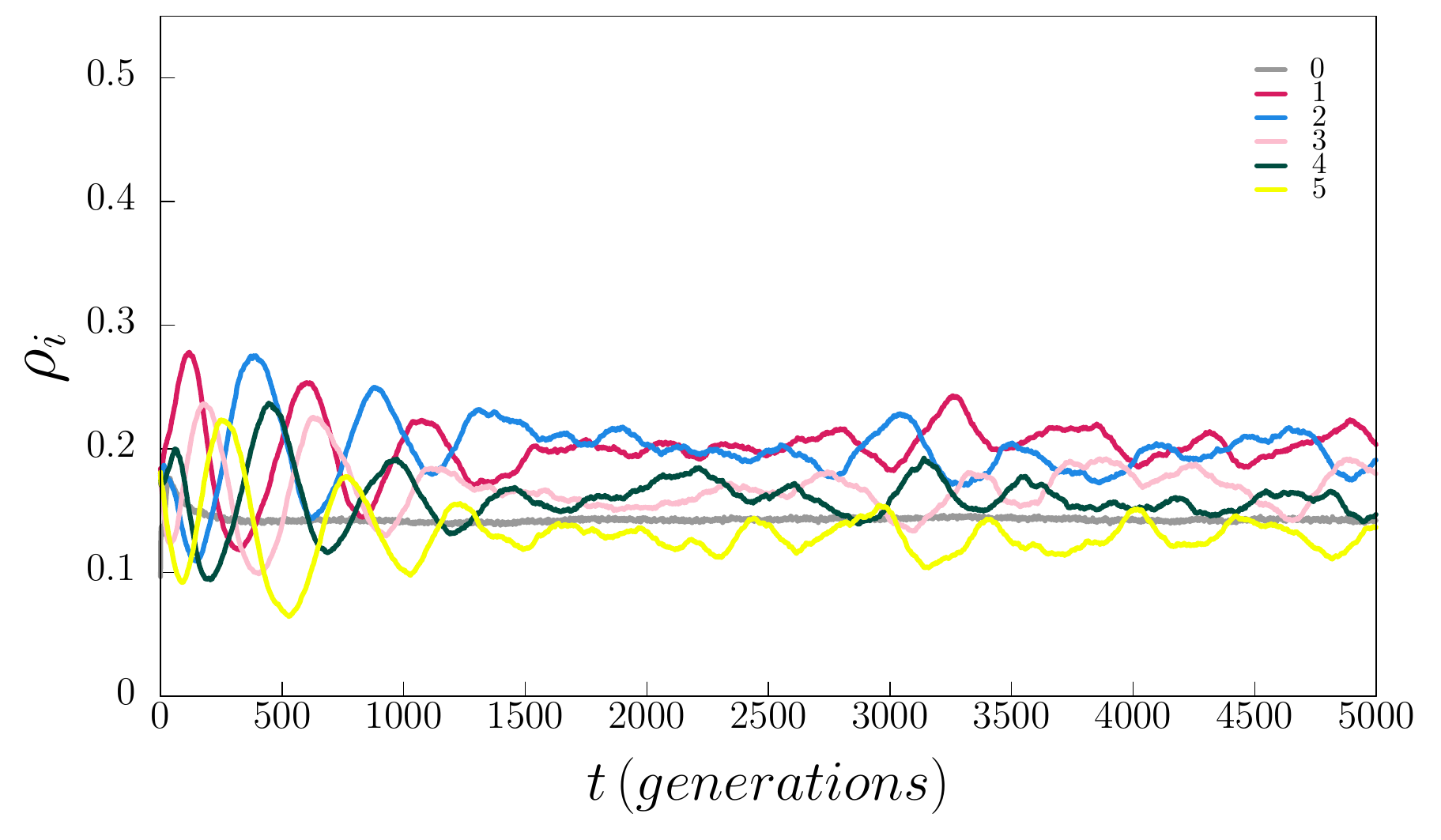}
        \caption{}\label{fig3b}
    \end{subfigure}
   \begin{subfigure}{.33\textwidth}
        \centering
        \includegraphics[width=57mm]{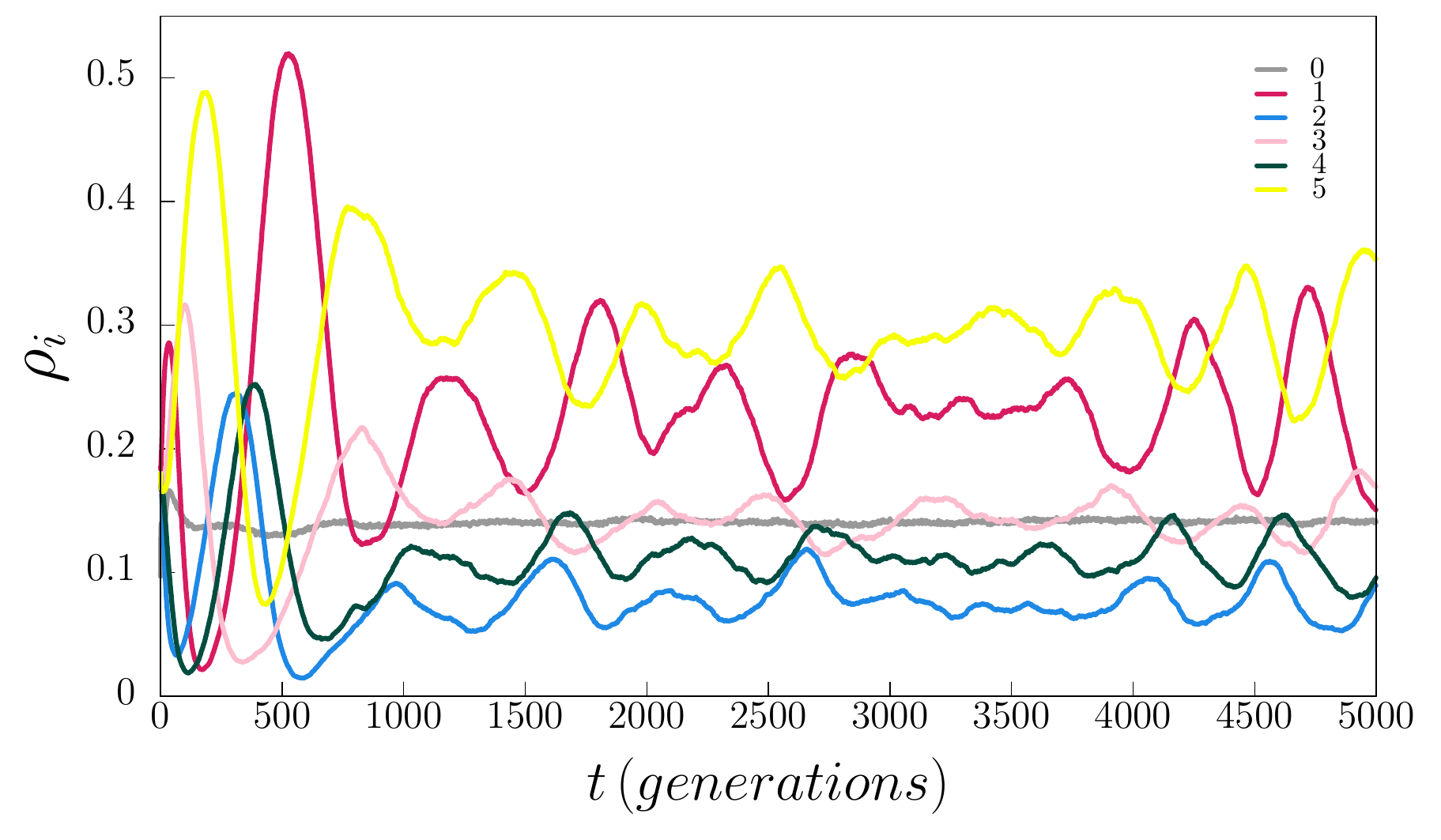}
        \caption{}\label{fig3c}
    \end{subfigure} 
\caption{Dynamics of the densities of species $i$ in the simulations shown in Fig.~\ref{fig2}. Figure a, b, and c shows the cyclic changes of the territorial occupation for species $i$ during the entire simulations for the standard model, Safeguard, and Social Distancing tactics, respectively. The colours follow the scheme in Fig.~\ref{fig1}.}
  \label{fig3}
\end{figure*}

\subsection{Organisms' Physical and Cognitive Ability}
In our model, organisms of one out of the species perform survival movement tactics. However, not all of them are physically or cognitively ready to move according to the species strategy. Moreover, the range of neighbourhood perception may vary, limiting the observation of the environment. Because of this, we define two parameters: 
\begin{itemize}
\item
Perception radius, $R$: the maximum Euclidian distance an individual can scan their vicinity, measured in lattice spacing, to identify the more attractive direction according to the collective strategy. 
\item
Conditioning factor, $\alpha$: a real parameter, with $0\,\leq\,\alpha\,\leq\,1$, representing the fraction of organisms conditioned to scan the vicinity and interpret the signals to accurately perform the behavioural strategy.
\end{itemize}
The standard model, where is represented by either $R=0$ or 
$\alpha = 0$, meaning that organisms cannot perceive or interpret the neighbourhood. On the opposite, the totality of the organisms follows the collective strategic movement for $\alpha = 1$.

\subsection{Self-preservation Movement Strategies}
Individuals of one out of the species can interpret the signals received from the environment to choose the best direction to move according to the survival movement tactics:  
\begin{enumerate}
\item 
Safeguard Strategy: an individual of species $i$ walks into the direction with the highest concentration of individuals of species $i-2$ \cite{Moura}. The aim is to avoid being killed by organisms of species $i-1$. In an epidemic scenario, there are two classes of Safeguard strategy:
\begin{itemize}
\item
Asymptomatic illness: healthy and sick organisms cannot be differentiated; thus, the social interaction may lead organisms to approach disease vectors.
\item
Symptomatic illness: organisms can discern what organisms are 
ill, keeping the social interaction exclusively with the healthy ones; thus, minimising the chances of being infected. The real parameter $\eta$, with $0 \leq \eta \leq 1$ indicates the percentage 
of incidence of symptomatic disease ($1-\eta$ are the fraction of sick organisms without symptoms).
\end{itemize}
\item
Social distancing Strategy: an individual of species $i$ moves towards the direction with more empty spaces. The main goal is to stay as far as possible from disease vectors.
\item
Combination of survival strategies: An individual od species $1$ combines Safeguard and Social Distancing strategies, with $\gamma$ indicating the priority to escape infection;  $\gamma$ is social distancing parameter, a real parameter, with $0 \leq \gamma \leq 1$, which represents the proportion of movements towards the direction with the higher density of empty spaces.
\end{enumerate}
In the standard model, no directional movement is performed, with all individuals of every species moving randomly. 

\subsection{Implementation of Strategic Directional Movement}
To implement the directional movement, the code proceeds the steps \cite{MENEZES2022101606}:
\begin{enumerate} 
\item 
implementing a circular area for an organism to observe the vicinity (a disc of radius $\mathcal{R}$, centred in the active individual); 
\item
separating the observation disc into four circular sectors in the directions of the nearest neighbour (the von Neumann neighbourhood defines the immediate vicinity); 
\item
counting the number of empty spaces and organisms of each species within each circular sector; organisms on the circular sector borders are assumed to be part of both circular sectors; 
\item
choosing the circular sector that contains the larger number of organisms of species $i-2$ ($h_{i-2}+s_{i-2}$) if the strategy executed is Safeguard and the disease is asymptomatic;
\item
selecting the circular sector that contains the larger number of healthy organisms of species $i-2$ ($h_{i-2}$) in the case of Safeguard tactic in an epidemic of symptomatic disease;
\item
choosing the circular sector with more empty spaces, whether the tactic is Social Distancing;
\item
in case of more the one direction being equally attractive, a draw between the tied directions is realised;
\item
switching positions of the active individual with the immediate neighbour in the direction of the selected circular sector. 
\end{enumerate}

\subsection{Spatial Autocorrelation Function}
\begin{figure}
 \centering
        \begin{subfigure}{.4\textwidth}
        \centering
        \includegraphics[width=75mm]{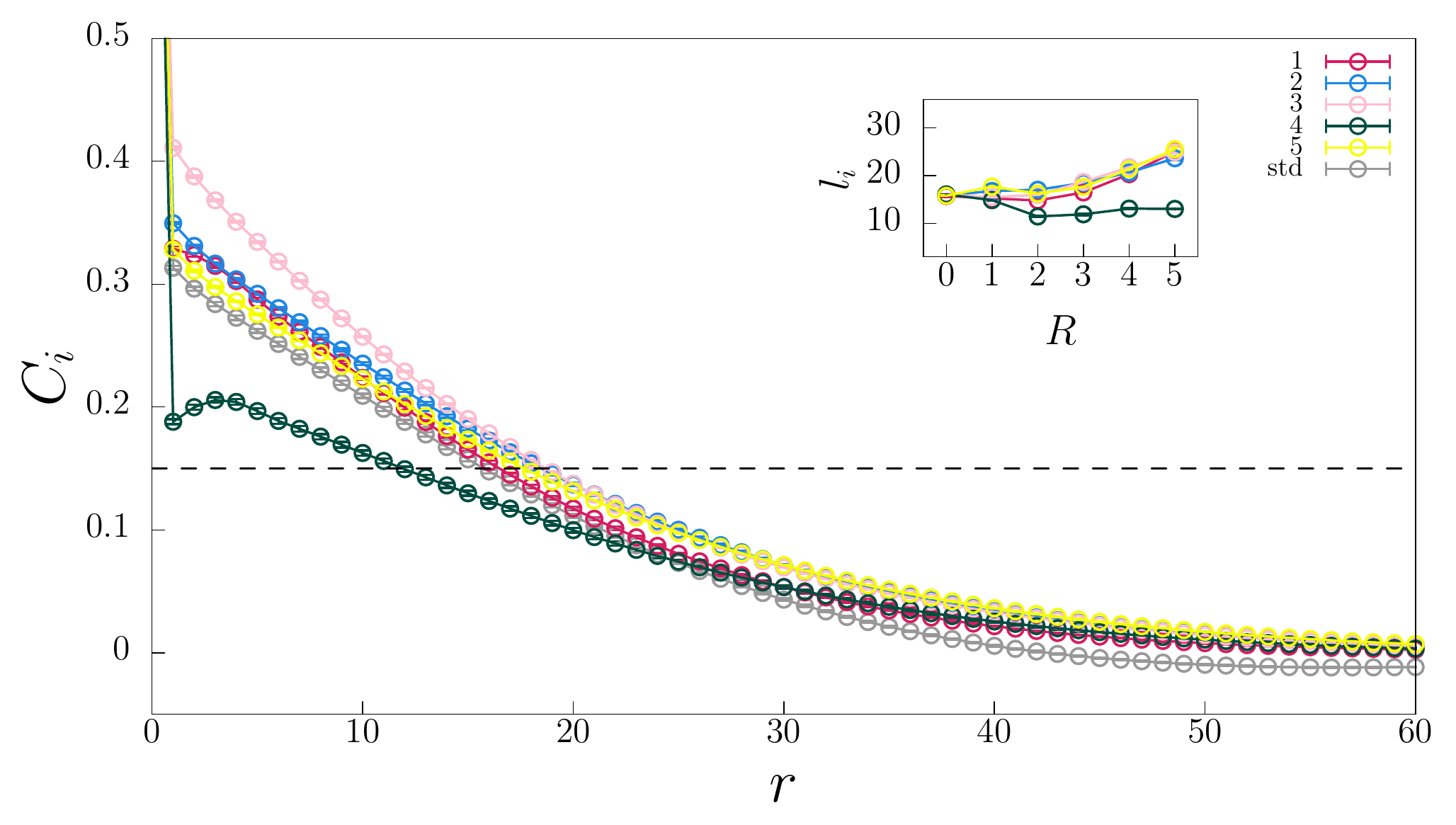}
        \caption{}\label{fig4a}
    \end{subfigure}\\
       \begin{subfigure}{.4\textwidth}
        \centering
        \includegraphics[width=75mm]{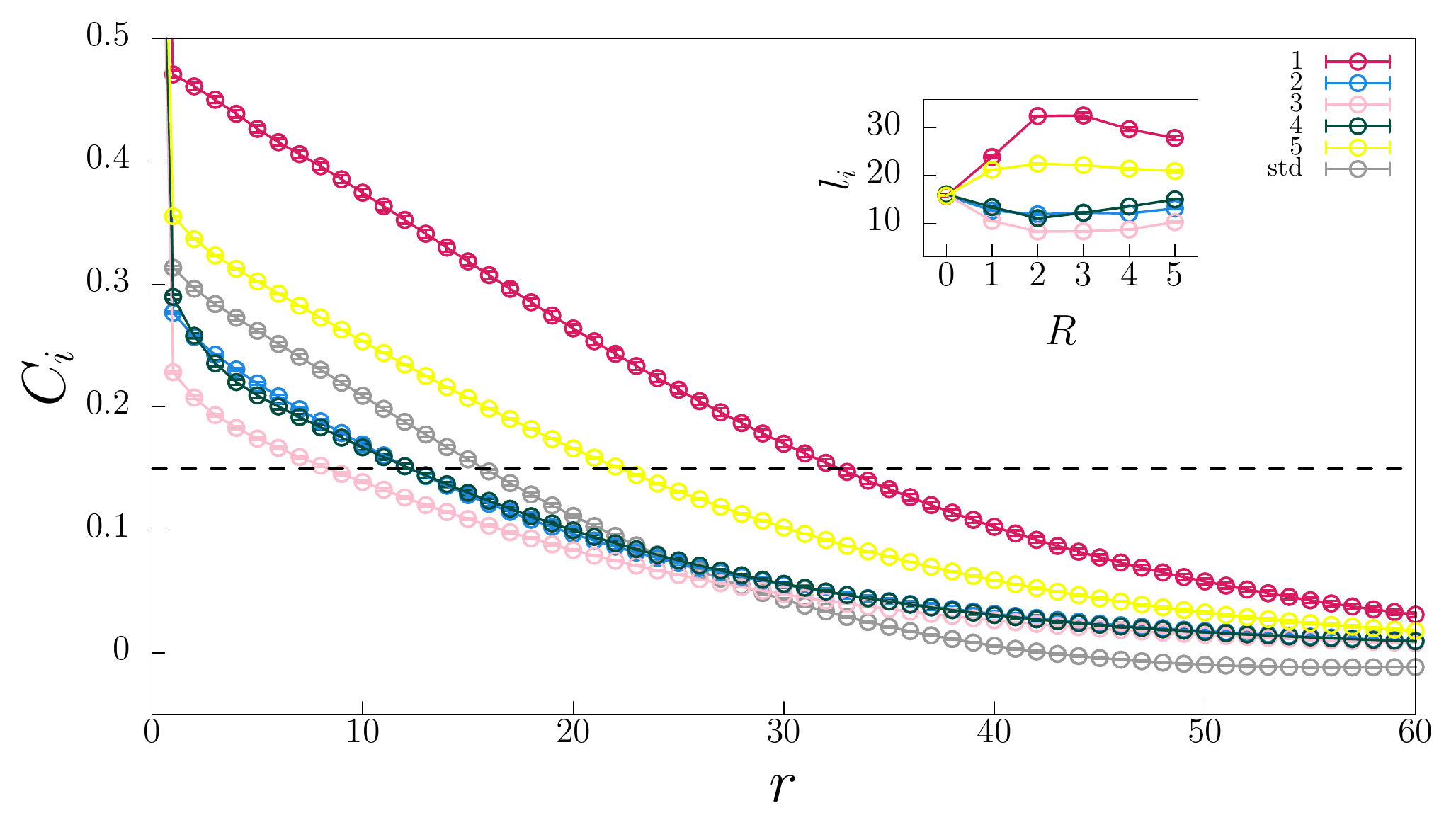}
        \caption{}\label{fig4b}
    \end{subfigure}\\
\caption{Spatial autocorrelation function and characteristic length of the typical areas occupied for species $i$. The mean autocorrelation functions for $R=3$ are depicted in Figures a and b for the Safeguard and Social Distancing strategies. The inset shows the variation of $l_i$ for various values of perception radius.}
  \label{fig4}
\end{figure}
\begin{figure*}
 \centering
    \begin{subfigure}{.33\textwidth}
    \centering
    \includegraphics[width=57mm]{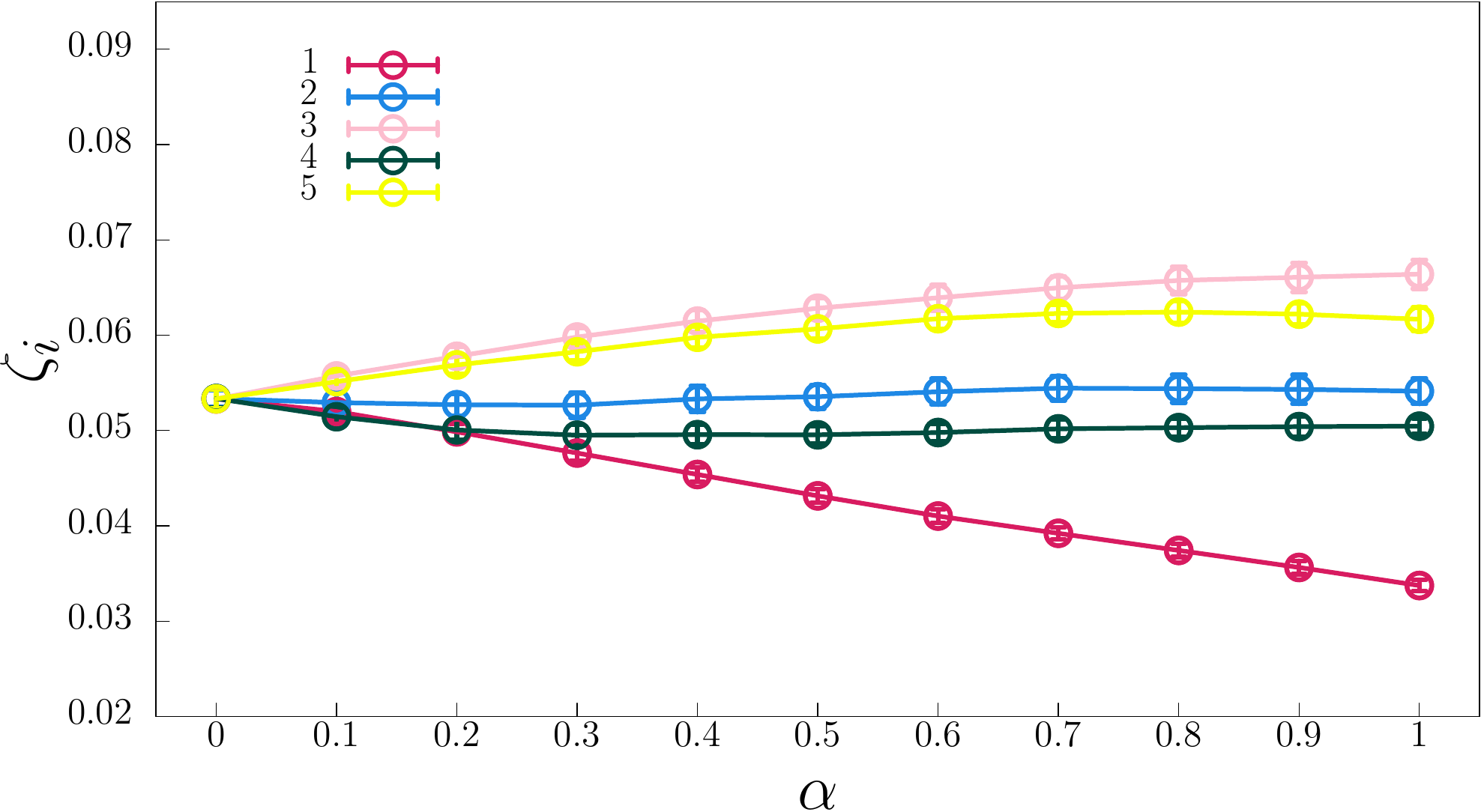}
    \caption{}\label{fig5a}
  \end{subfigure}
    \begin{subfigure}{.33\textwidth}
    \centering
    \includegraphics[width=57mm]{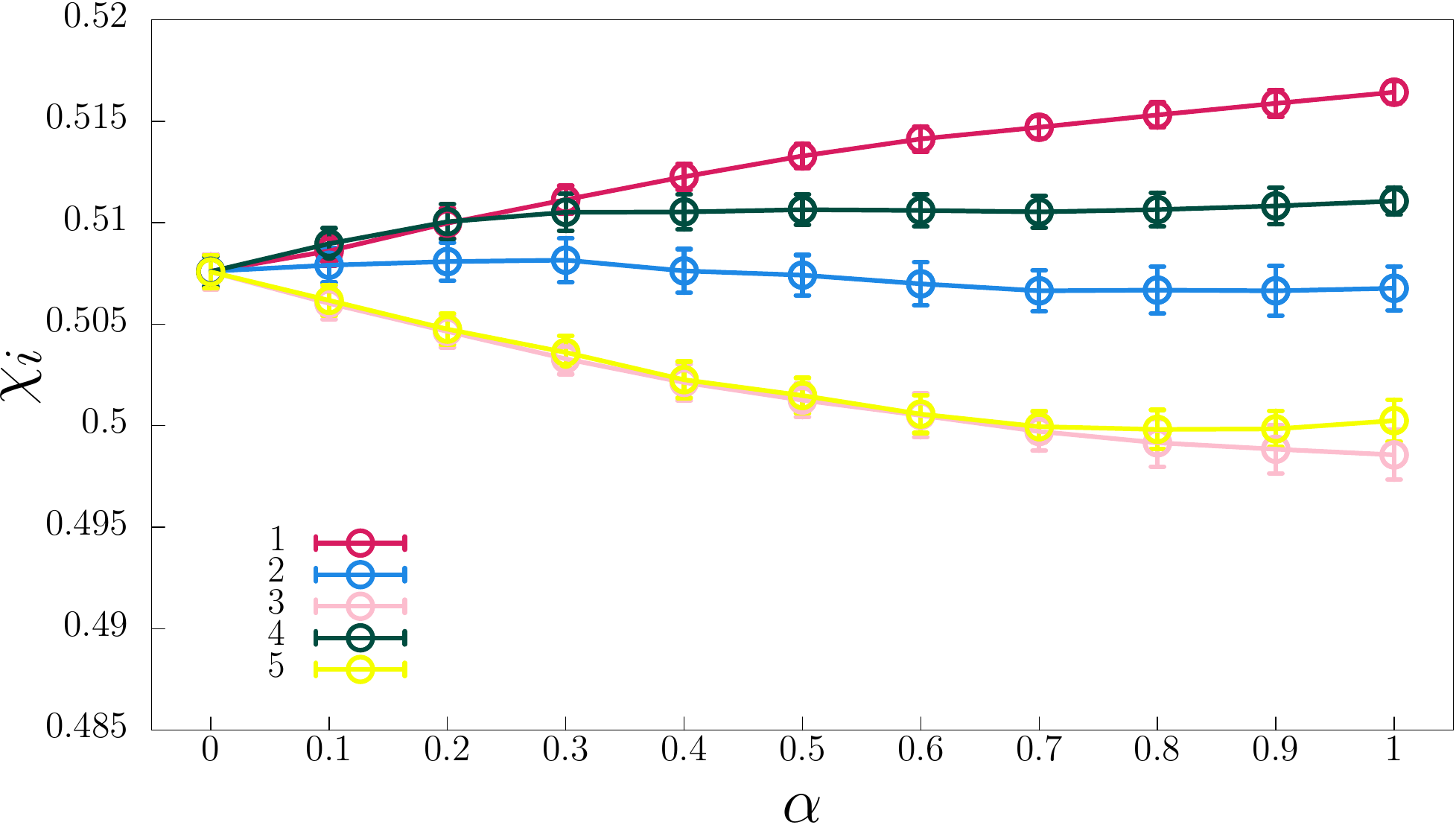}
    \caption{}\label{fig5b}
  \end{subfigure}
      \begin{subfigure}{.33\textwidth}
    \centering
    \includegraphics[width=57mm]{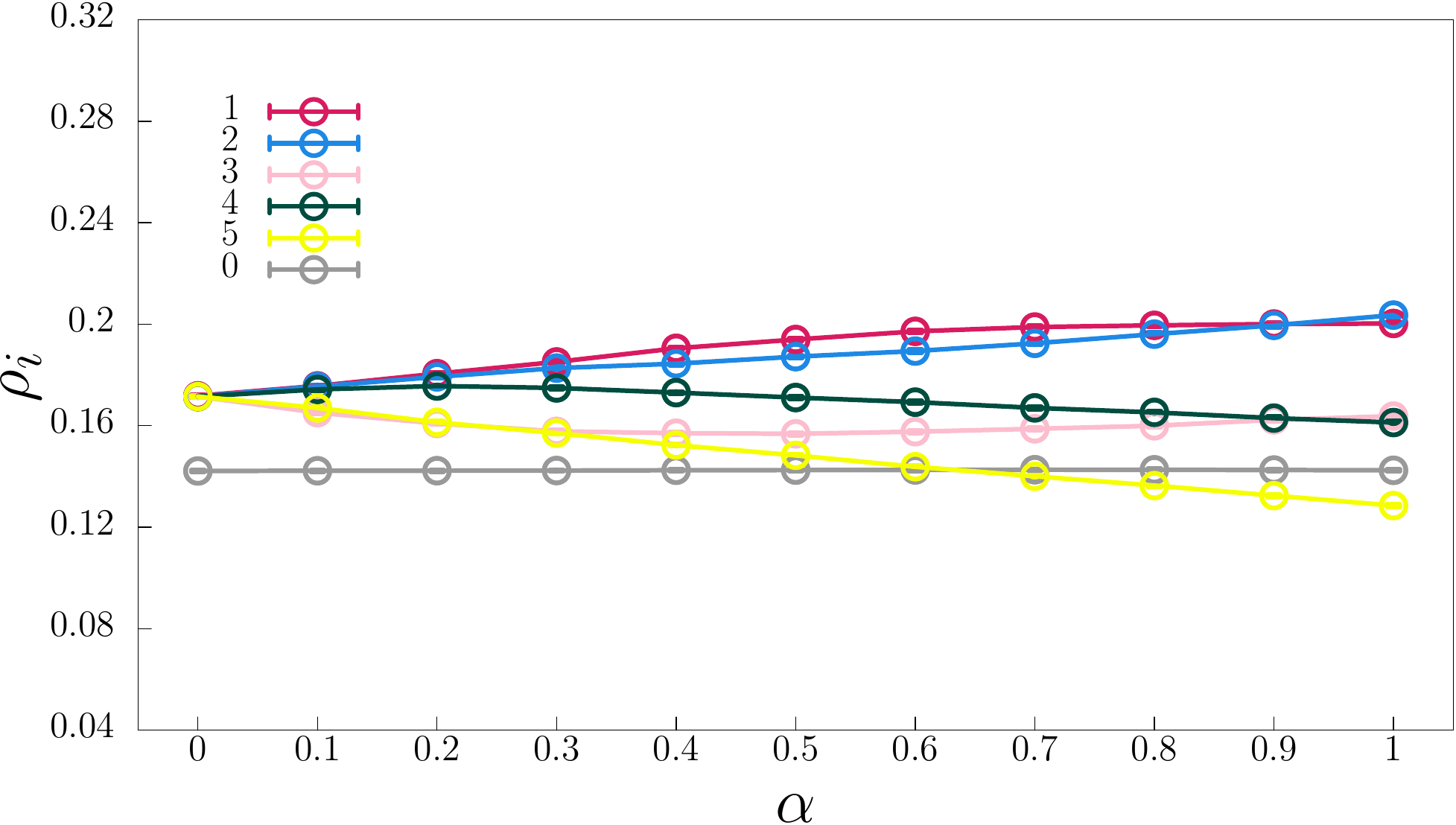}
    \caption{}\label{fig5c}
  \end{subfigure}
  \\
   \begin{subfigure}{.33\textwidth}
    \centering
    \includegraphics[width=57mm]{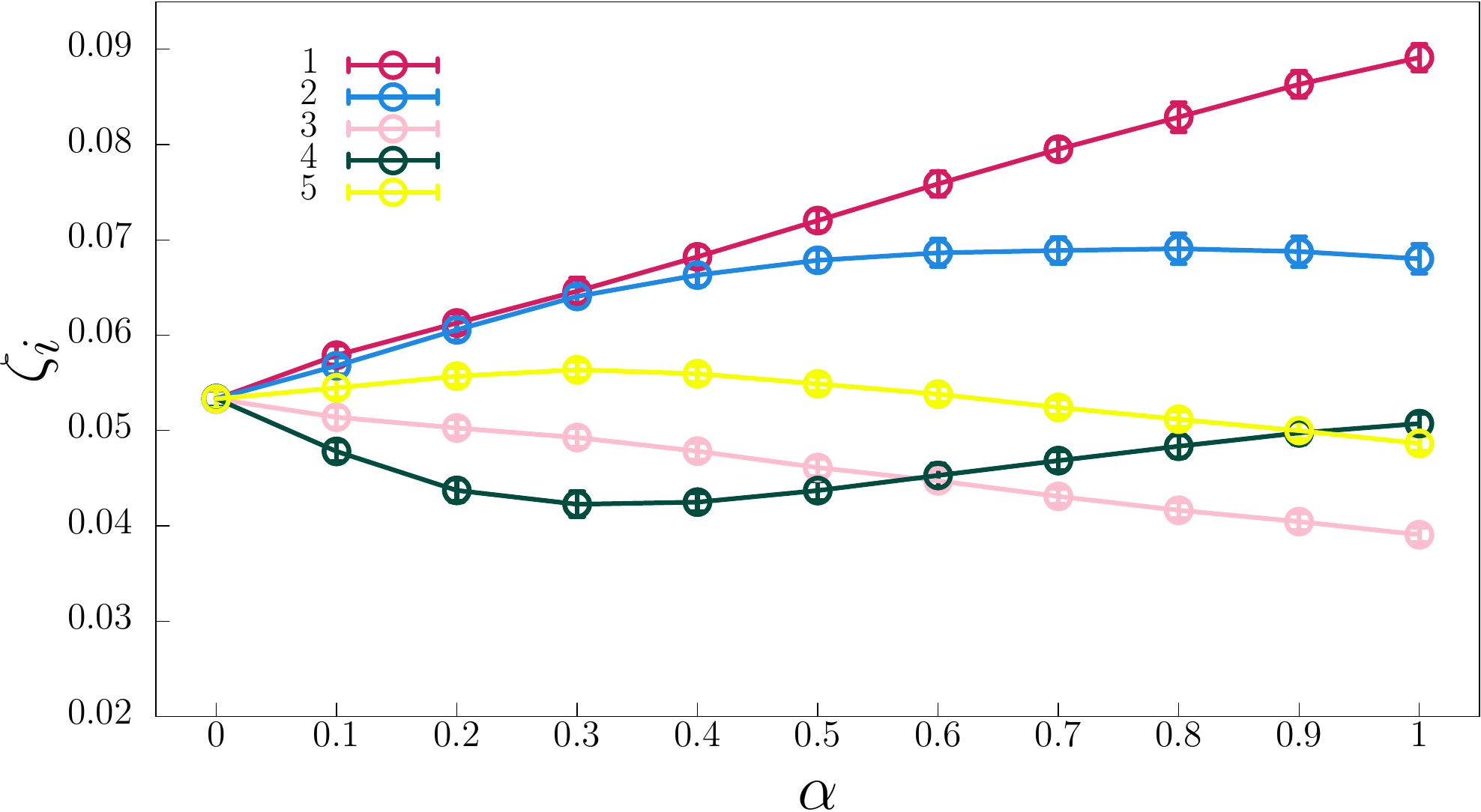}
    \caption{}\label{fig5d}
  \end{subfigure}
    \begin{subfigure}{.33\textwidth}
    \centering
    \includegraphics[width=57mm]{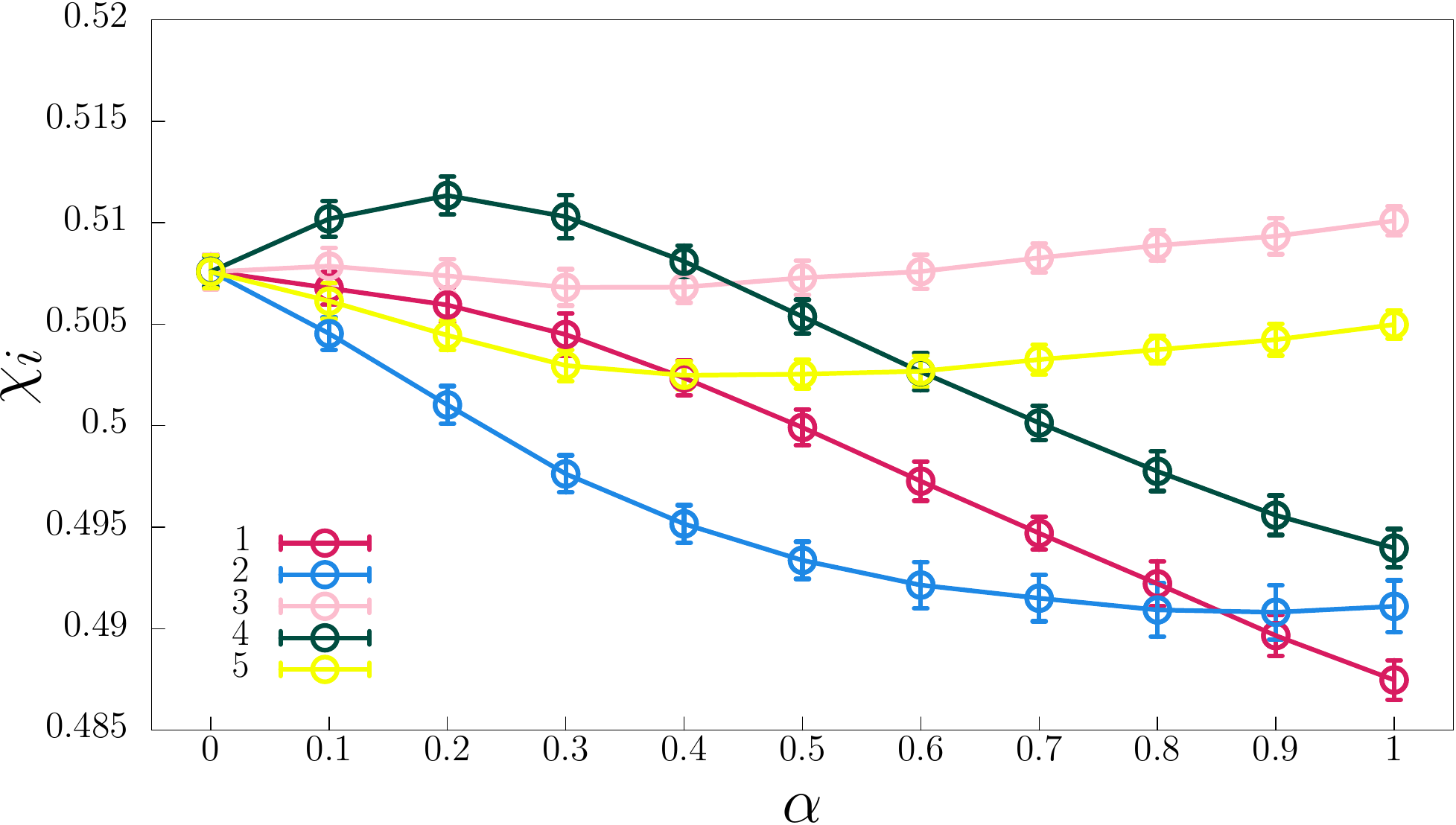}
    \caption{}\label{fig5e}
  \end{subfigure}
      \begin{subfigure}{.33\textwidth}
    \centering
    \includegraphics[width=57mm]{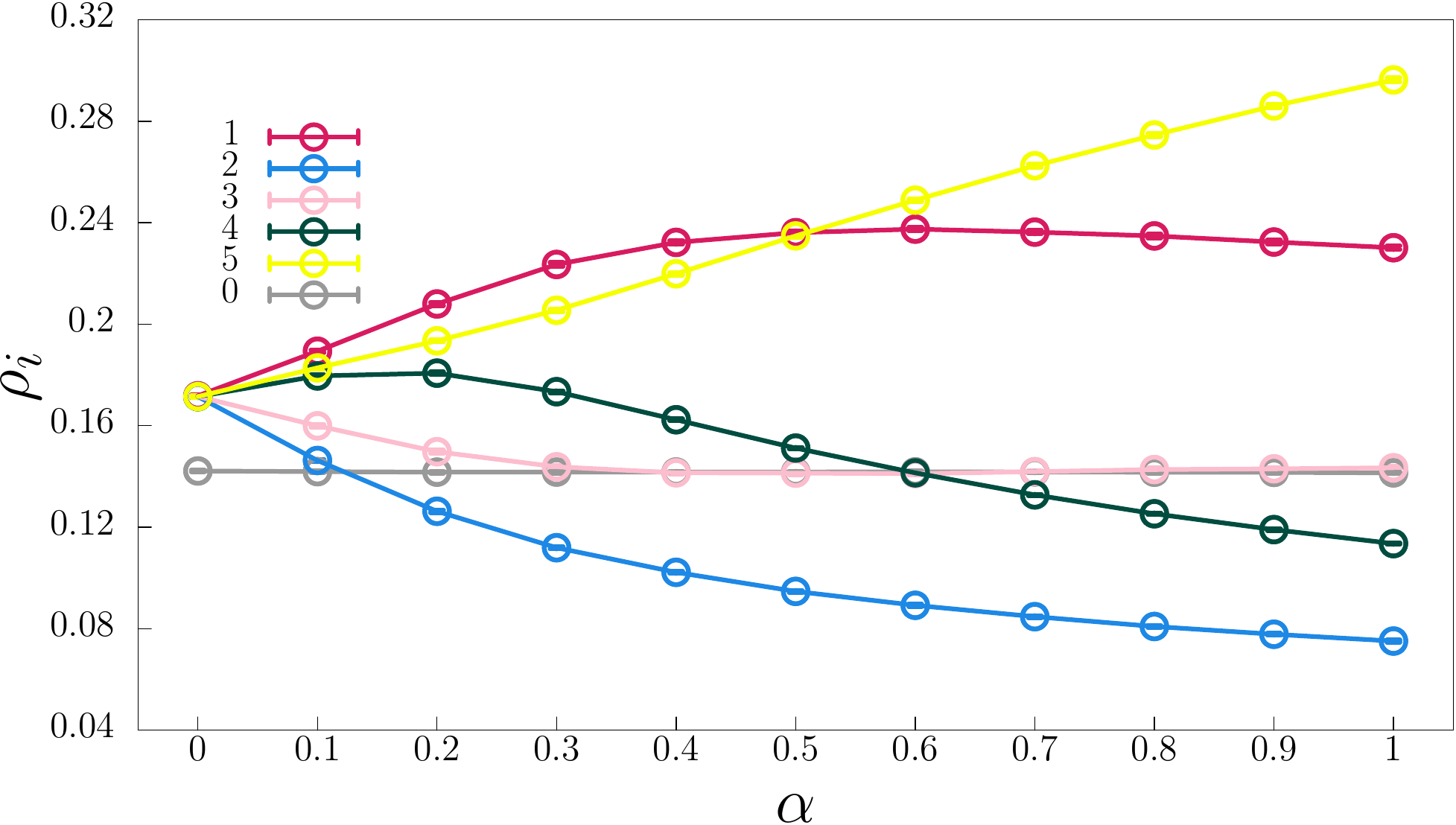}
    \caption{}\label{fig5f}
  \end{subfigure}
  \caption{Selection risk, predation risk, densities of species as functions of the conditioning factor. Figures a, b, and c show the results for the Safeguard
tactic, whereas the outcomes for the Social Distancing strategy are shown in Figures d, e, and f. The standard deviation is indicated by error bars.}
 \label{fig5}
\end{figure*}

To calculate the characteristic length defining the scale of spatial domains occupied by each species, we first calculate the spatial autocorrelation function $C_i(r)$, with $i=1,2,3,4,5$, in terms of radial coordinate $r$, where $r=|\vec{r}|=x+y$ is the Manhatan distance between $(x,y)$ and $(0,0)$. 
We define the function $\phi_i(\vec{r})$ to describe the position $\vec{r}$ in the lattice occupied by individuals  of species $i$. Calculating the mean value $\langle\phi_i\rangle$, we find the Fourier transform
\begin{equation}
\varphi_i(\vec{\kappa}) = \mathcal{F}\,\{\phi_i(\vec{r})-\langle\phi_i\rangle\},
\end{equation}
that gives the spectral densities
\begin{equation}
S_i(\vec{k}) = \sum_{k_x, k_y}\,\varphi_i(\vec{\kappa}).
\end{equation}

The autocorrelation function is found by employing the normalised inverse Fourier transform
\begin{equation}
C_i(\vec{r}') = \frac{\mathcal{F}^{-1}\{S_i(\vec{k})\}}{C(0)}.
\end{equation}
The spatial autocorrelation function for species $i$ as 
a function of the radial coordinate $r$ is then written as
\begin{equation}
C_i(r') = \sum_{|\vec{r}'|=x+y} \frac{C_i(\vec{r}')}{min\left[2N-(x+y+1), (x+y+1)\right]}.
\end{equation}
Finally, once the spatial autocorrelation is known, the
typical size of the spatial domains of organisms of species $i$ is calculated by 
assuming the threshold $C_i(l_i)=0.15$, where $l_i$ is the characteristic length scale for spatial domains of species $i$.

\subsection{Selection and Infection Risks}

To explore how an organism's safety is impacted by the directional movement tactic, we calculate two risks:
\begin{itemize}
\item
Selection Risk, $\zeta_i(t)$: the probability of an organism of species $i$ (irrespective of the health condition) being eliminated by an individual of species $i-1$ at time $t$. 
The implementation of the selection risk follows the algorithm:
i) counting the total number of individuals of species $i$ at the beginning of each generation; ii) computing how many individuals of species $i$ are selected (killed by individuals of species $i-1$) during the generation; iii) calculating the selection risk, $\zeta_i$, with $i=1,2,3,4,5$, as the ratio between the number of consumed individuals and the initial amount.
\item
Infection Risk, $\chi_i(t)$: the probability of a healthy organism of species $i$ being infected by an ill individual of any species at time $t$. The infection risk is implemented as follows:
i) calculating the total number of healthy individuals of species $i$ at the beginning of each generation; ii) computing the number of individuals of species $i$ infected during the generation; 
iii) calculating the infection risk, $\chi_i$, with $i=1,2,3,4,5$, as the ratio between the number of infected individuals and the initial number of healthy organisms.
\end{itemize}

\subsection{Model parameters}
 
All outcomes presented throughout this paper were obtained from simulations in square lattices with $500^2$ sites, running for a timespan of $5000$ generations, assuming the set of probabilities: $s_h = r_h = m_h = 1/3$, $s_s = r_s = m_s = w = 5/22$, and $c = d = 1/22$. We have verified that our conclusions also hold for other sets of probabilities. Moreover, all statistical analysis whose results appear in Figs.~\ref{fig4} to ~\ref{fig7} were realised by performing series of $100$ simulations, starting from different initial conditions.

\section{Spatial patterns}
\label{sec3}

We first investigated how the self-defence movement strategies impact the pattern formation process. Running single simulations, we assumed
all organisms of species $1$ to be conditioned to move strategically, $\alpha=1.0$. The random initial condition depicted in the snapshot of Fig.~\ref{fig2a} were used in three simulations: i) Simulation A: all organisms of every species move randomly (standard model); ii) Simulation B: individuals of species $1$ move towards the largest group of organisms of species $4$ (Safeguard), whereas organisms of other species move randomly; ii) Simulation C: organisms of species $1$ move towards the direction with more empty spaces (Social Distancing), while individuals of other species walk aleatorily.
The results are depicted in Figs.~\ref{fig2b}, ~\ref{fig2c}, and ~\ref{fig2d}, respectively, where the colours follow the scheme in Fig.~\ref{fig1}: red, blue, pink, green, and yellow dots show individuals of species $1$, $2$, $3$, $4$, and $5$; empty spaces appear in white. 
Additionally, Figs.~~\ref{fig3a}, ~\ref{fig3b}, and ~\ref{fig3c} shows the dynamics of densities of species in the simulations A, B, and C, respectively.

Simulation A shows that all organisms are equally vulnerable to being infected and killed by selection in the standard model. The outcomes show the appearance of spiral waves whose arms are mostly occupied by organisms of two species; namely, $\{1,3\}$, $\{1,4\}$, $\{2,4\}$, $\{2,5\}$, and $\{3,5\}$. Fig.~\ref{fig2a} shows
that empty spaces are equally distributed throughout the grid as a result of death from the epidemic disease. Figure \ref{fig3a} shows the cyclic species predominance resulting from the rock-paper-scissors rules, with the average densities being the same for every species.

However, the symmetric spiral wave formation is no longer present if species $1$ move directionally. 
Figure~\ref{fig2b} reveal what happens in the case of Safeguard strategy,  where organisms of species $1$ scan their neighbourhood to walk in the direction with more guards (individuals of species $4$): 
species $2$ can multiply since fewer individuals of species $1$ access them, resulting in a decline of the population of species $3$. This allows species $4$ to control a more significant fraction of the lattice, providing plenty of refuge to species $1$. Therefore, the fraction of territory occupied by species $1$ increases because fewer organisms are eliminated. However, the Safeguard tactic helps the disease spread, causing more organisms' death.
Thus, as depicted in Fig.~\ref{fig3b}, the safeguard movement tactic does not benefit species $1$ as happens in the absence of an epidemic \cite{Moura}.

On the other hand, whether organisms execute the Social Distancing tactic, species $1$ profits doubly: reproduces more (because moving to areas of the higher density of empty spaces) and reduces the chances of contamination by possible infected organisms. The spatial patterns in Fig.~\ref{fig2d} show that the social distancing strategy also benefits species $5$, whose organism can conquer territory occupied by species $1$. Thus, our findings demonstrated that despite waiving the helpful protection offered by species $4$, the Social Distancing tactic is determining to the population growth of species $1$ in 
an epidemic scenario, as depicted in Fig.~\ref{fig3c}.

\section{Characteristic Length Scales}
\label{sec4}
We compute the typical size of species groups by calculating the spatial autocorrelation function of species $i$ analysing the spatial configuration at $t=5000$ generations of a group of $100$ simulations. The simulations were repeated for various $R$ to clarify the role of the perception radius on the spatial species segregation.
Figures \ref{fig4a} and \ref{fig4b} shows the autocorrelation function as a function of the radial coordinate for Safeguard and Social Distancing strategies, respectively, for $R=3$ - the colours follow the scheme of Fig.~\ref{fig1}. 
The average autocorrelation function in the standard model is the same for every species, as depicted by a grey line in both figures. 
The inset figures show the characteristic length, $l_i$, with $R=0$ representing the standard model.

For the Safeguard tactic, the typic group size is approximately the same for every species, except for species $4$, which decreases compared to the standard model. Moreover, the directional movement strategy's effects on species segregation are more prominent for long-range perception. On the other hand, the impact of the species agglomeration is more significant in the case of Social Distancing: there is an enlargement of the characteristic length of species $1$ and $4$, with $l_1>l_4$, irrespective of the perception radius. In contrast, the typical regions inhabited mainly by individuals of species $2$, $3$ and $5$ diminishes, with $l_2$ being the shortest, independent of $R$. The outcomes also reveal that Social Distancing more significantly influences the spatial patterns for intermediary values of $R$.

\section{The role of the conditioning factor}
\label{sec5}
We now focus on the role of the conditioning factor in the survival movement strategies in an epidemic. The mean value of the selection risk $\zeta_i$, infection risk $\xi_i$, and species densities $\rho_i$ are depicted in Figs.~\ref{fig5a}, ~\ref{fig5b}, and ~\ref{fig5c} for Safeguard, and Figs.~\ref{fig5d}, ~\ref{fig5e}, and ~\ref{fig5f}, for Social Distancing tactics, respectively. The standard deviation is shown in the error bars; the colours follow the scheme in Fig.~\ref{fig1}. The results for the standard model, where all organisms move randomly, are represented by $\alpha=0$.

Overall, the outcomes show that the more organisms participate in the movement strategy, the more accentuated the impact in the population dynamics. In the case of Safeguard tactics, as $\alpha$ grows, 
selection risk of species $1$ drops, 
(Fig.~\ref{fig5a}), but the the chances of becoming ill rises (Fig.~\ref{fig5b}).
Despite the increase of the infection risk, the balance is positive in terms of population size when compared with the standard model (Fig.~\ref{fig5c}). Besides, species $4$ also profits from the strategic movement of safeguarding of species $1$ with a reduced selection risk.
However,  the losses provoked by the exposure to infection due to the proximity to organisms of species $1$ outweigh the benefits of reducing the selection risk. Thus, the density of species $4$ decreases as $\alpha$ grows. Besides species 1, the only one that profits with the increase in the controlled territorial area (compared to the standard model) is species 2, which predominates in the cyclic game in the limit $\alpha=1$.

In contrast, as the fraction organisms of species $1$ apt to perform Social Distancing grows, $\zeta_1$ rises (Fig.~\ref{fig5d}), while $\xi_1$ drops (Fig.~\ref{fig5e}). Overall, the strategic movement is more advantageous if all organisms move directionally, resulting in the maximum density of species $1$. Although for $\alpha \leq 0.2$, species $4$ benefits with an increase of the territorial control, the species that profits more is species $5$, whose spatial density sharply rises as $\alpha$ increases, being preponderant for $\alpha \geq 0.5$.
\begin{figure}[t]
\centering
    \begin{subfigure}{.4\textwidth}
        \centering
        \includegraphics[width=73mm]{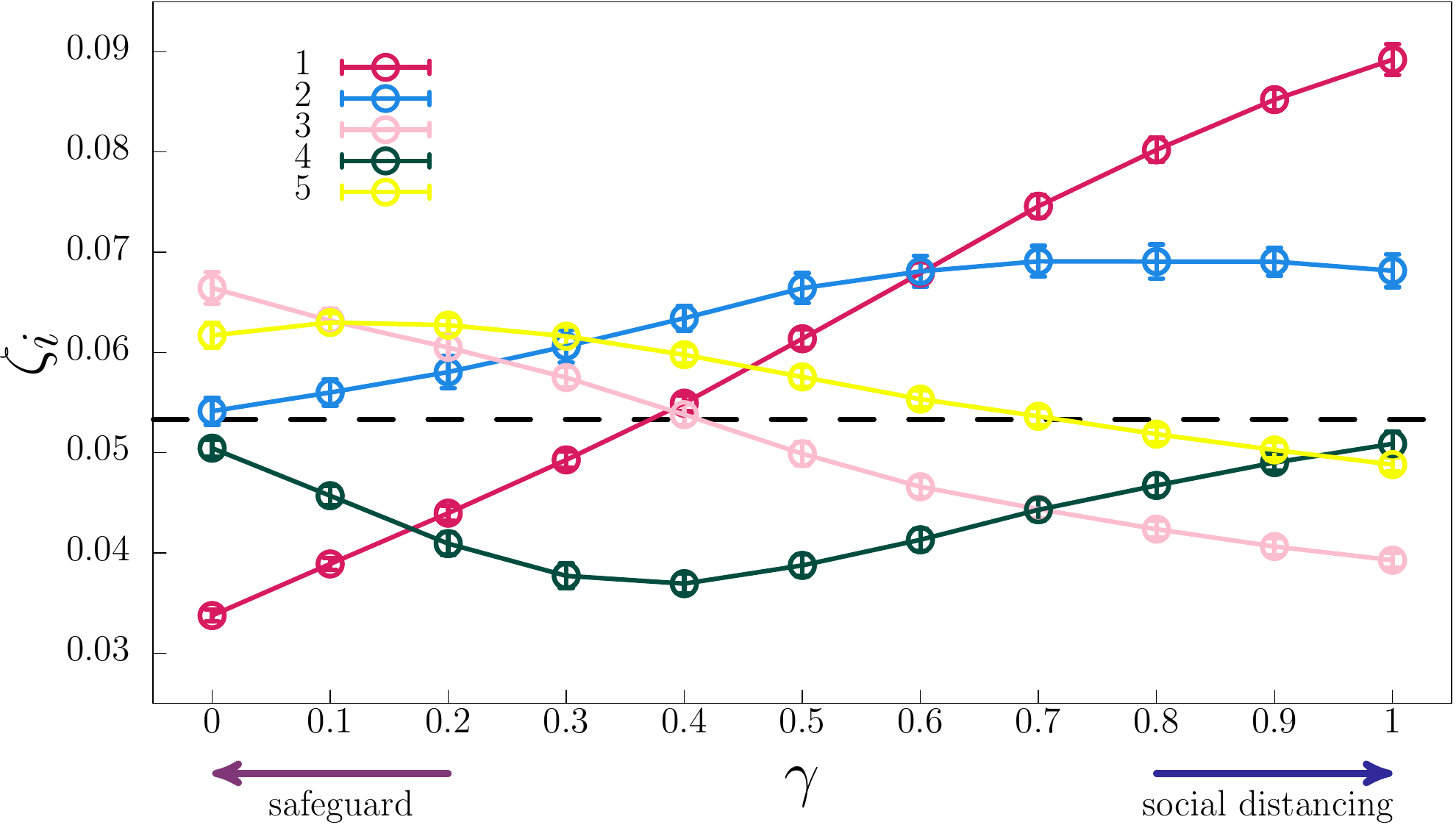}
        \caption{}\label{fig6a}
    \end{subfigure} %
       \begin{subfigure}{.4\textwidth}
        \centering
        \includegraphics[width=73mm]{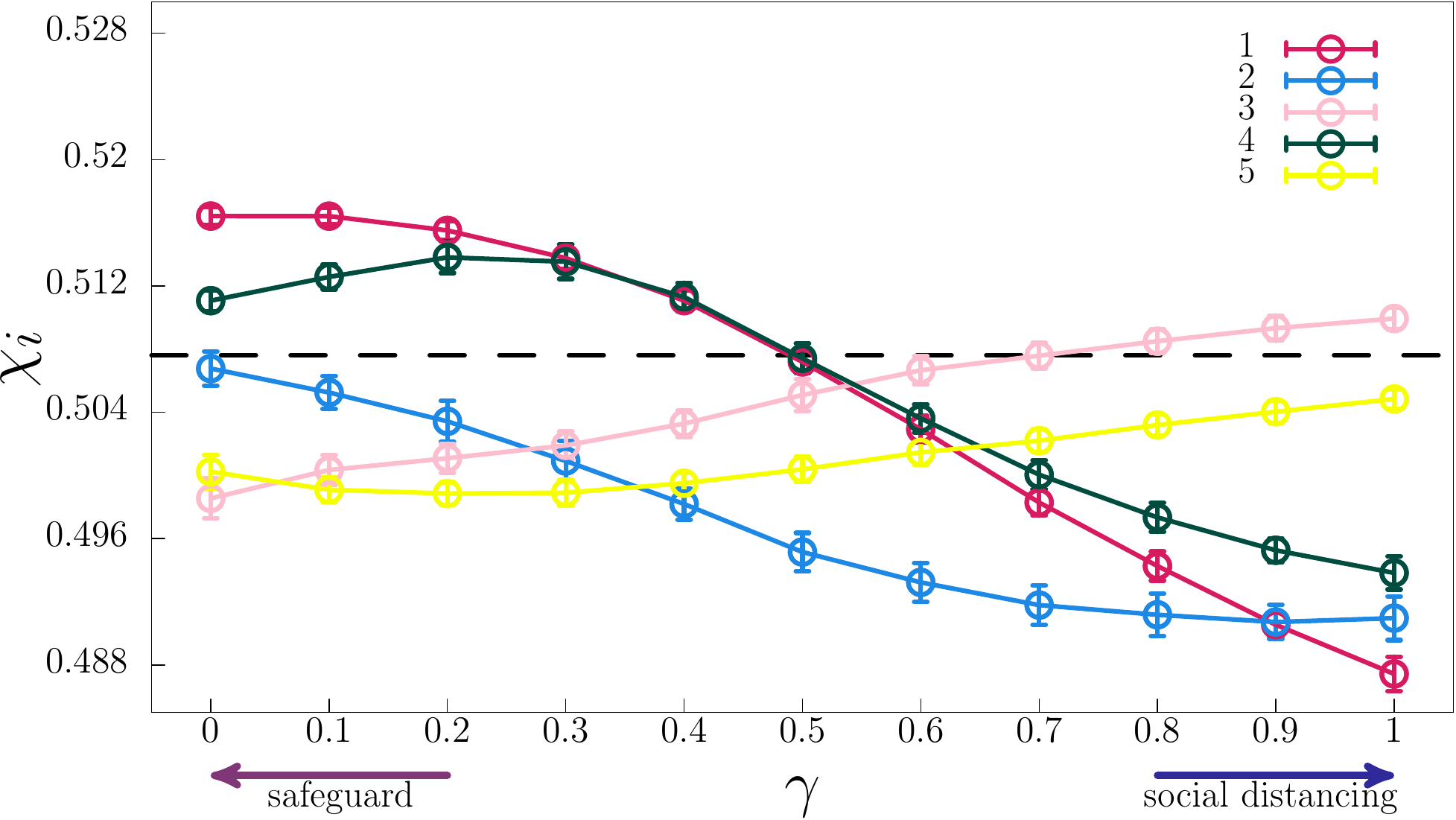}
        \caption{}\label{fig6b}
    \end{subfigure} %
           \begin{subfigure}{.4\textwidth}
        \centering
        \includegraphics[width=73mm]{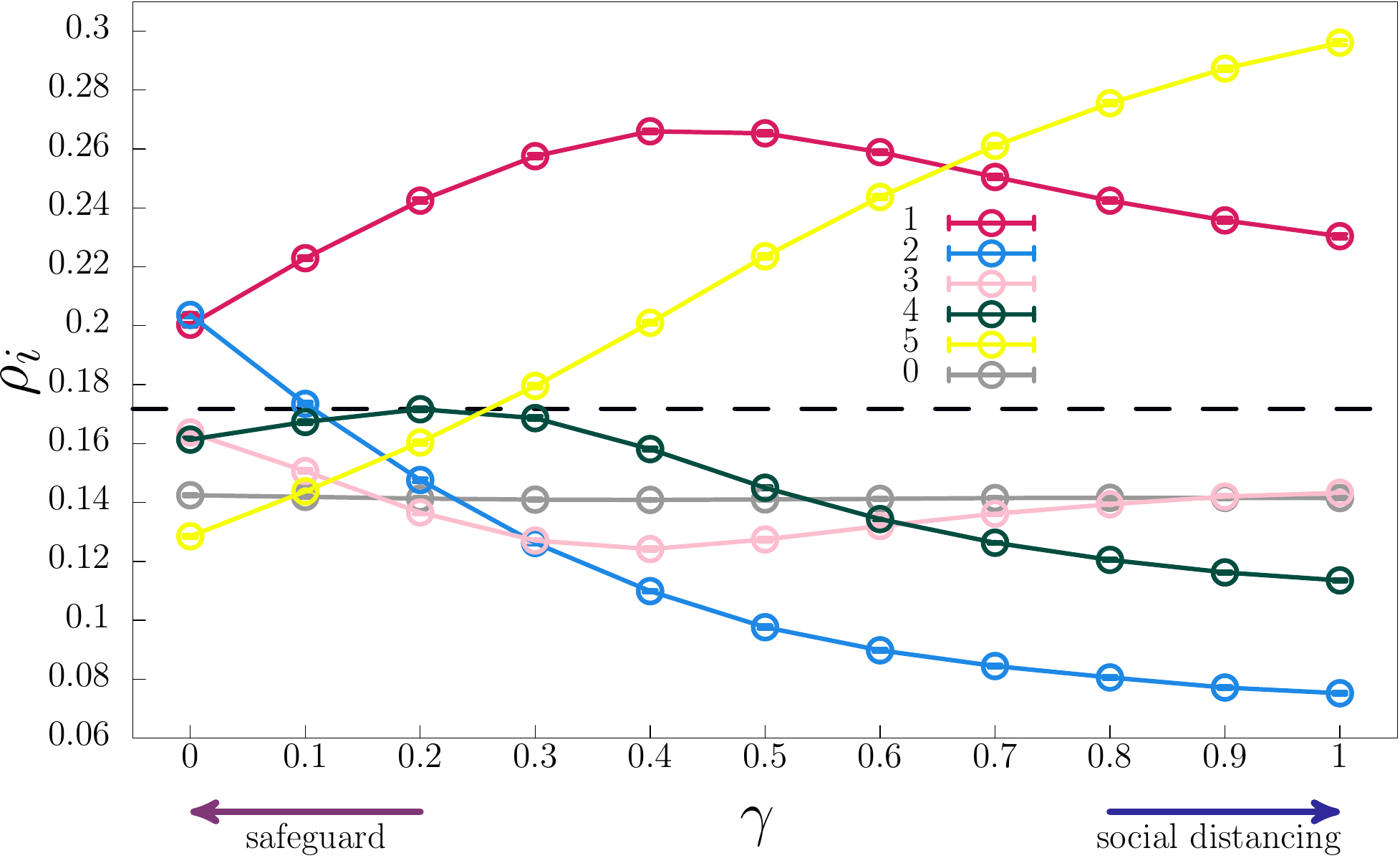}
        \caption{}\label{fig6c}
    \end{subfigure} %
    \caption{Selection risk (Figure a), predation risk (Figure b), densities of species (Figure c) as functions of the social distancing factor. 
The black horizontal dashed lines indicate the selection risk, infection risk, and densities of species in the standard model.}
  \label{fig6}
\end{figure}
\section{Combination the self-preservation movement tactics}
\label{sec6}
In the previous sections, we studied the effects of each self-preservation strategy in pattern formation and population dynamics. Now, we explore the case where organisms of species $1$ combine both behavioural tactics to maximise territorial control in the cyclic game system.
Figs.~\ref{fig6a}, \ref{fig6b}, and \ref{fig6c} depict $\zeta_i$, $\chi_i$, and $\rho_i$ as a function of $\gamma$, the social distancing factor; the colours follows the scheme in Fig.~\ref{fig1}. 
The horizontal dashed black line indicates the value assumed by $\zeta_i$, $\xi_i$, and $\rho_i$ in the standard model, where organisms of every species move randomly. 

The outcomes show that maximum protection against one of the death causes (disease or attack of an enemy) is achieved if the respective movement strategy is prioritised. 
Namely, as $\gamma$ increases, more susceptible to selection and less vulnerable to viral infection an individual of species $1$ becomes. 
However, Fig.~\ref{fig6c} reveals that combining both strategies is more advantageous.
Our findings show that if $40\%$ of the movements are guided by the Social Distancing and $60\%$ by Safeguard tactic, $\rho_1$ is maximum. 
Therefore, $\gamma=0.4$ is the combination that leads species $1$ to maximise territorial control. In summary, in comparison with the standard model - whose results are indicated by the black dashed line: i) the population of species $1$ increases for any combination of the survival strategies; ii) species $2$ benefits from the behaviour of organisms of species $1$ if $\gamma \leq 0.1$; iii) similar to species $2$ but is smaller scale, species $3$ also gains territory if $\gamma \leq 0.1$; iv) spatial density of species $4$ grows for $\gamma \leq 0.3$; v) species $5$ is the more benefited from the Social Distancing tactic performed by species $1$, which brings a significant population growth if $\gamma \geq 0.3$.
\section{Survival movement strategies in a symptomatic disease epidemic}
\label{sec7}

\begin{figure}[t]
\centering
    \begin{subfigure}{.4\textwidth}
        \centering
        \includegraphics[width=73mm]{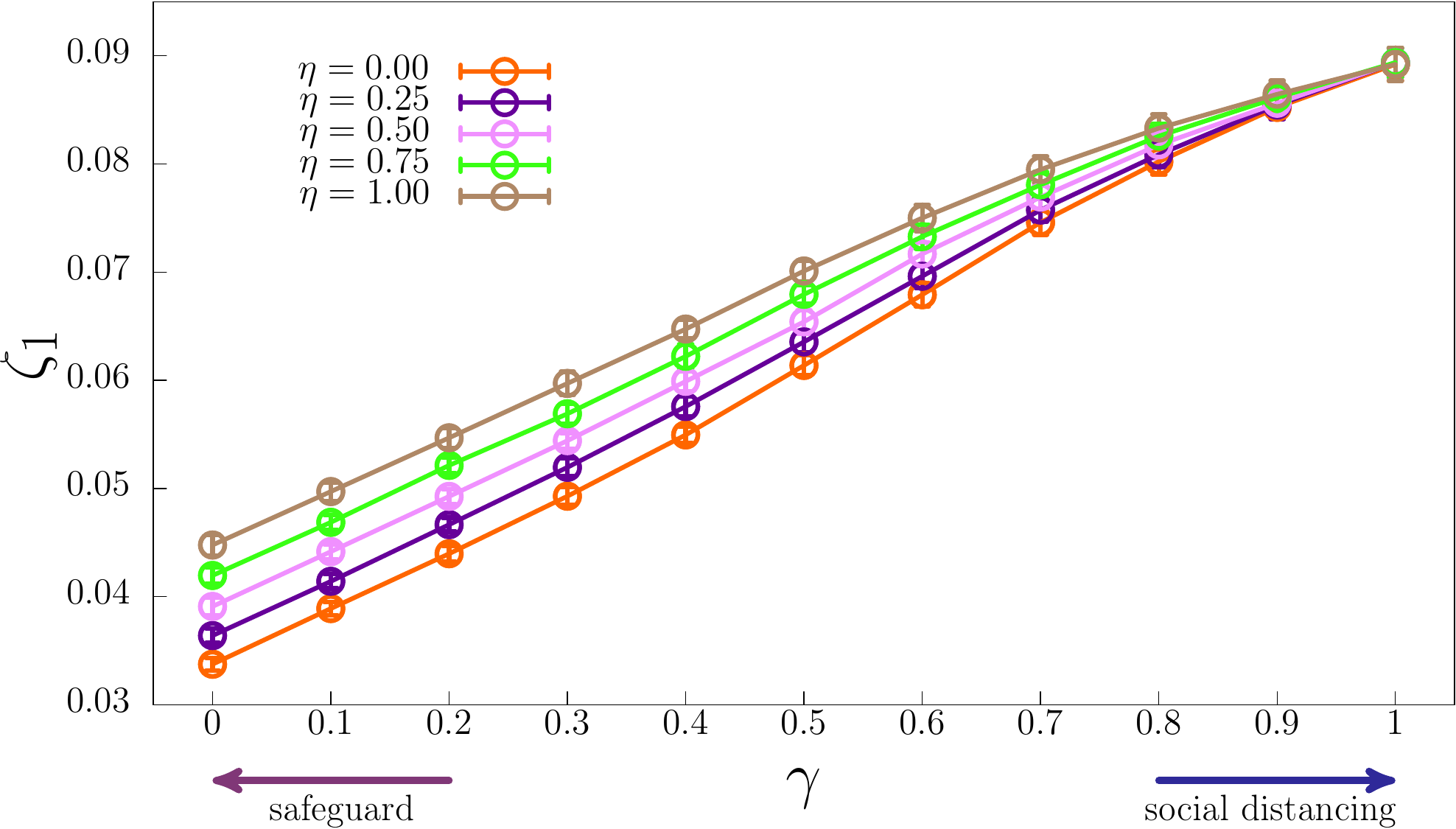}
        \caption{}\label{fig7a}
    \end{subfigure} %
       \begin{subfigure}{.4\textwidth}
        \centering
        \includegraphics[width=73mm]{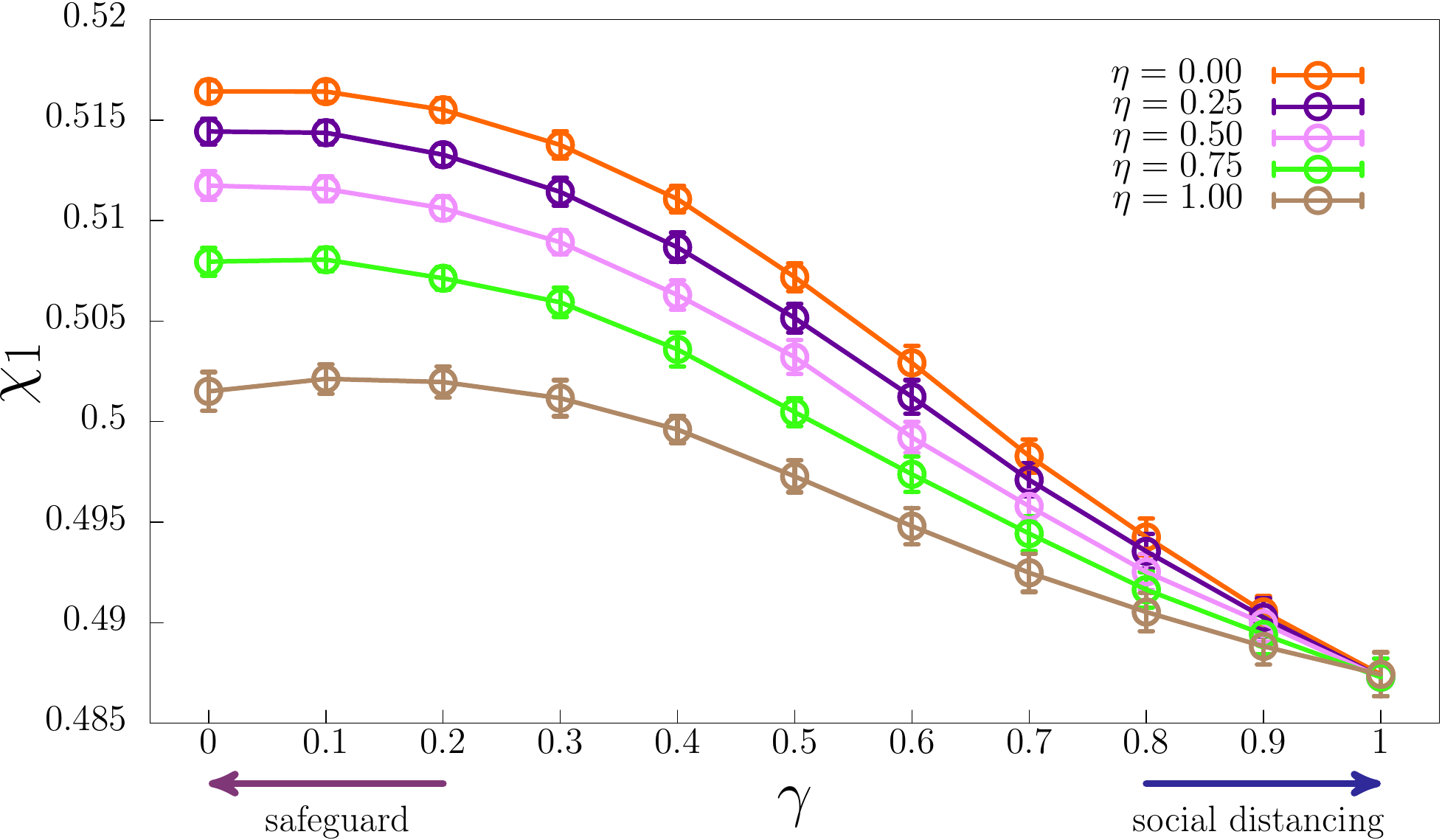}
        \caption{}\label{fig7b}
    \end{subfigure} %
           \begin{subfigure}{.4\textwidth}
        \centering
        \includegraphics[width=73mm]{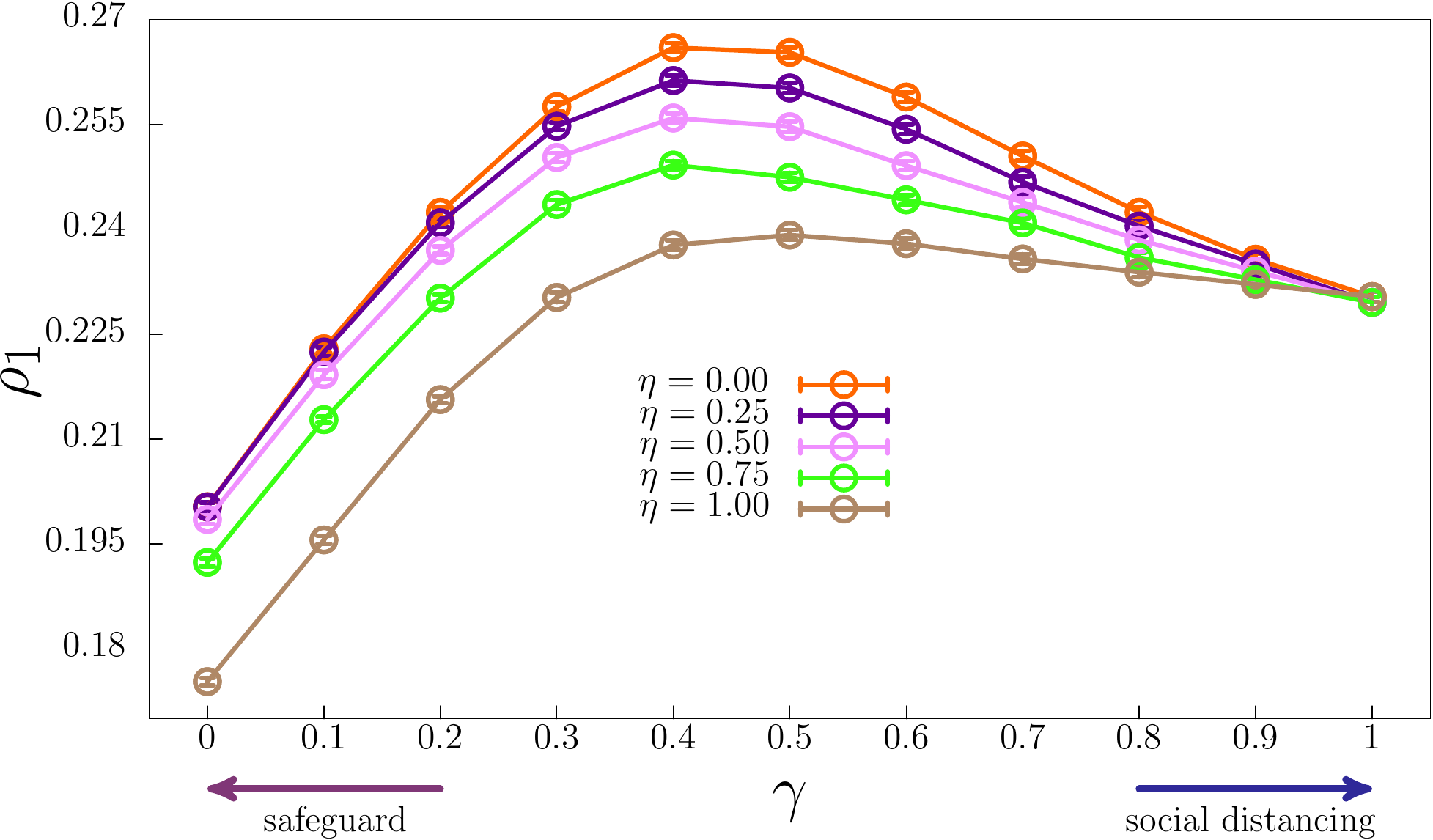}
        \caption{}\label{fig7c}
    \end{subfigure} %
    \caption{Selection risk (Figure a), predation risk (Figure b), densities of species (Figure c) of species $i$ in terms of the social distancing factor for various symptomatic factor.
Orange, dark purple, light purple, green, brown lines depict the outcomes for epidemic where the percentage of sick organisms with symptoms is $0\%$, $25\%$, $50\%$, $75\%$, $100\%$, respectively. The standard deviation is indicated by error bars.}
  \label{fig7}
\end{figure}

Finally, we explore a scenario where organisms of species $1$ can distinguish between healthy and ill individuals. This means that whenever an individual executes the Safeguard strategy, it disregards sick organisms of species $4$, thus moving towards the direction of more healthy individuals. Furthermore, we assume that ill organisms are identified because of the symptoms. Therefore, we explore a variety of scenarios where not all sick organisms have symptoms, thus limiting the capacity of individuals of species $1$ avoiding approaching them.
Figs.~\ref{fig7a} and ~\ref{fig7b} show the outcomes for $\eta=0.0$ (orange line), $\eta=0.25$ (dark purple line), $\eta=0.5$ (light purple line), $\eta=0.75$ (green line), and $\eta=1.0$ (brown line), where $\eta$ indicates the percentage of sick individuas with symptoms, for Safeguard and Social Distancing strategies, respectively. 

We found that the more significant the proportion of ill organisms with symptoms, the less risky for individuals of organisms of species $1$ to be infected when performing the Safeguard tactic. Moreover, if all ill individuals have symptoms ($\eta=1.0$), the infection risk is minimal, irrespective of the fraction of individuals performing the Safeguard strategy. 
On the other hand, Fig.~\ref{fig7b} shows that the larger $\eta$ is, the riskier is to individuals of species $1$ being selected by dominant species. This happens because avoiding approaching infected refuges means losing protection that infected organisms of species $4$ offer against selection. We conclude that the Safeguard tactic is more profitable in the case of asymptomatic disease ($\eta=0.0)$. 

Furthermore, Fig.~\ref{fig7c} shows that the mean value of $\rho_1$ is maximum for $\gamma=0.4$ except for the case of all ill organisms having symptoms ($\eta=1.0$), where $\rho_i$ is maximum for $\gamma=0.5$. 
Moreover, if organisms exclusively move according to Safeguard strategy, the fraction of the grid controlled by species $1$ is maximum if not more than $25\%$ of ill individuals have symptoms. In the case of only Social Distancing being performed, the percentage of ill organisms with symptoms does not matter because social approaching is avoided to minimise the chances of interacting with an ill asymptomatic ill individual.

\section{Discussion and Conclusions}
\label{sec8}

We studied a stochastic cyclic model with five species in an epidemic scenario. The disease is transmitted from person to person, irrespective of the species. Using an individual-based algorithm, we simulated the case where organisms of one out of the species perform survival movement strategies to protect themselves against: i) being selected by an individual of the dominant species; ii) being infected by an ill organism.
Combining the Safeguard and the Social Distancing tactics allows the organism to move towards areas with low population density (low density of disease vectors) and high concentration of guards (high density of enemies of organism's enemies). 
Our modelling considers that the behavioural strategies' performance depends on the individuals' physical and cognitive capacities to scan the environment and interpret the signals to choose the best direction to move.

Our findings reveal that the movement strategies change the pattern formation, leading to unevenness in the territory occupation by the species. Remarkably, the species whose organisms practice self-preservation tactics benefit from population growth compared with the standard model. Furthermore, our outcomes reveal that when only one of the self-defence tactics is executed, Social Distancing is more profitable, giving the species the control of a more significant fraction of the lattice. Nevertheless, the best result is obtained if attention is divided on self-protecting from the two threats to its existence: selection and disease infection. The best profit is achieved if organisms focus on Social Distancing $40\%$ of the times they move - for the set of parameters assumed in our simulations. Under these conditions, the average fraction of the territory controlled by the species executing the behavioural movement strategy grows more than $50\%$ compared to the standard model.

In an epidemic of symptomatic disease, where identifying infected organisms is possible, the Safeguard strategy leads to different results according to the proportion of ill individuals with symptoms. Namely,
the more significant the proportion of symptomatic sick individuals
the less dangerous the Safeguard strategy is - the infection risk drops. However, symptoms bring a side effect since when bypassing infected organisms, a number of refuges against selection are disregarded, reducing the efficiency of the Safeguard strategy. Therefore, the spatial density of the species combining survival strategies declines as the fraction of ill individuals with symptoms grows, irrespective of the percentage of displacements dedicated to each particular tactic.

Our conclusions can be generalised to the case of a disease whose severity varies due to, for example, a virus mutation. If the disease becomes more transmissible or lethal, Social Distancing plays
a more vital role in organisms' survival. Specifically, the fraction of movements moving towards the direction with more empty spaces must increase. On the contrary, whether organisms cured of the disease retain an immunological memory, which guarantees permanent or temporary protection against reinfection, the role of Social Distancing declines. In this case, the best territorial control is reached if organisms prioritise the Safeguard strategy, maximising the benefits of self-protection against selection in the cyclic game.

Our results show the role of behaviour in responding strategically when 
facing environmental death risks.
Our conclusions can be generalised for more complex scenarios with more species competing for space in a generalised rock-paper-scissors game. The outcomes may be helpful to the understanding of the balance between diverse organisms' strategies in ecosystems facing dynamic epidemic scenarios.

\section*{Acknowledgments}
We thank CNPq, ECT, Fapern, and IBED for financial and technical support.
\bibliographystyle{elsarticle-num}
\bibliography{ref}

\end{document}